\documentstyle[twoside]{article}
\catcode`\@=11
\long\def\@makefntext#1{
\protect\noindent \hbox to 3.2pt {\hskip-.9pt
$^{{\eightrm\@thefnmark}}$\hfil}#1\hfill}		

\def\thefootnote{\fnsymbol{footnote}}
\def\@makefnmark{\hbox to 0pt{$^{\@thefnmark}$\hss}}	

\def\ps@myheadings{\let\@mkboth\@gobbletwo
\def\@oddhead{\hbox{}
\rightmark\hfil\eightrm\thepage}
\def\@oddfoot{}\def\@evenhead{\eightrm\thepage\hfil
\leftmark\hbox{}}\def\@evenfoot{}
\def\sectionmark##1{}\def\subsectionmark##1{}}
\oddsidemargin=\evensidemargin
\renewcommand{\thefootnote}{\fnsymbol{footnote}}

\newcounter{sectionc}\newcounter{subsectionc}\newcounter{subsubsectionc}
\renewcommand{\section}[1] {\vspace{12pt}\addtocounter{sectionc}{1}
\setcounter{subsectionc}{0}\setcounter{subsubsectionc}{0}\noindent
	{\tenbf\thesectionc. #1}\par\vspace{5pt}}
\renewcommand{\subsection}[1] {\vspace{12pt}\addtocounter{subsectionc}{1}
	\setcounter{subsubsectionc}{0}\noindent
	{\bf\thesectionc.\thesubsectionc. {\kern1pt \bfit #1}}\par\vspace{5pt}}
\renewcommand{\subsubsection}[1] {\vspace{12pt}\addtocounter{subsubsectionc}{1}
	\noindent{\tenrm\thesectionc.\thesubsectionc.\thesubsubsectionc.
	{\kern1pt \tenit #1}}\par\vspace{5pt}}
\newcommand{\nonumsection}[1] {\vspace{12pt}\noindent{\tenbf #1}
	\par\vspace{5pt}}
\newcommand{\textlineskip}{\baselineskip=13pt}
\newcommand{\smalllineskip}{\baselineskip=10pt}
\def\eightcirc{
\begin{picture}(0,0)
\put(4.4,1.8){\circle{6.5}}
\end{picture}}
\def\eightcopyright{\eightcirc\kern2.7pt\hbox{\eightrm c}}
\newcommand{\copyrightheading}[1]
	{\vspace*{-2.5cm}\smalllineskip{\flushleft
	{\footnotesize International Journal of Modern Physics A, #1}\\
	{\footnotesize $\eightcopyright$\, World Scientific Publishing
	 Company}\\
	 }}
\newcommand{\pub}[1]{{\begin{center}\footnotesize\smalllineskip
	Received #1\\
	\end{center}
	}}

\def\abstracts#1#2#3{{
	\centering{\begin{minipage}{4.5in}\baselineskip=10pt\footnotesize
	\parindent=0pt #1\par
	\parindent=15pt #2\par
	\parindent=15pt #3
	\end{minipage}}\par}}

\newcommand{\bibit}{\nineit}

\renewenvironment{thebibliography}[1]
	{\frenchspacing
	 \ninerm\baselineskip=11pt
	 \begin{list}{\arabic{enumi}.}
	{\usecounter{enumi}\setlength{\parsep}{0pt}
	 \setlength{\leftmargin 12.7pt}{\rightmargin 0pt} 
	 \setlength{\itemsep}{0pt} \settowidth
	{\labelwidth}{#1.}\sloppy}}{\end{list}}
\newcounter{itemlistc}
\newcounter{romanlistc}
\newcounter{alphlistc}
\newcounter{arabiclistc}
\newenvironment{itemlist}
    	{\setcounter{itemlistc}{0}
	 \begin{list}{$\bullet$}
	{\usecounter{itemlistc}
	 \setlength{\parsep}{0pt}
	 \setlength{\itemsep}{0pt}}}{\end{list}}
\newenvironment{romanlist}
	{\setcounter{romanlistc}{0}
	 \begin{list}{$($\roman{romanlistc}$)$}
	{\usecounter{romanlistc}
	 \setlength{\parsep}{0pt}
	 \setlength{\itemsep}{0pt}}}{\end{list}}

\def\@citex[#1]#2{\if@filesw\immediate\write\@auxout
	{\string\citation{#2}}\fi
\def\@citea{}\@cite{\@for\@citeb:=#2\do
	{\@citea\def\@citea{,}\@ifundefined
	{b@\@citeb}{{\bf ?}\@warning
	{Citation `\@citeb' on page \thepage \space undefined}}
	{\csname b@\@citeb\endcsname}}}{#1}}
\newif\if@cghi
\def\cite{\@cghitrue\@ifnextchar [{\@tempswatrue
	\@citex}{\@tempswafalse\@citex[]}}
\def\citelow{\@cghifalse\@ifnextchar [{\@tempswatrue
	\@citex}{\@tempswafalse\@citex[]}}
\def\@cite#1#2{{$\null^{#1}$\if@tempswa\typeout
	{IJCGA warning: optional citation argument
	ignored: `#2'} \fi}}

\def\pmb#1{\setbox0=\hbox{#1}
	\kern-.025em\copy0\kern-\wd0
	\kern.05em\copy0\kern-\wd0
	\kern-.025em\raise.0433em\box0}

\def\fnm#1{$^{\mbox{\scriptsize #1}}$}
\def\fnt#1#2{\footnotetext{\kern-.3em
	{$^{\mbox{\scriptsize #1}}$}{#2}}}
\def\fpage#1{\begingroup
\voffset=.3in
\thispagestyle{empty}\begin{table}[b]\centerline{\footnotesize #1}
	\end{table}\endgroup}
\def\runninghead#1#2{\pagestyle{myheadings}
\markboth{{\protect\footnotesize\it{\quad #1}}\hfill}
{\hfill{\protect\footnotesize\it{#2\quad}}}}
\font\tenrm=cmr10
\font\tenit=cmti10
\font\tenbf=cmbx10
\font\bfit=cmbxti10 at 10pt
\font\ninerm=cmr9
\font\nineit=cmti9

\font\eightrm=cmr8

\begin{document}

\runninghead{Quantum Geometrodynamics of the Bianchi-IX Model in Extended
Phase Space}{Quantum Geometrodynamics of the Bianchi-IX Model in Extended
Phase Space}

\normalsize\textlineskip
\thispagestyle{empty}
\setcounter{page}{1}

\copyrightheading{}			

\vspace*{0.88truein}

\fpage{1}
\centerline{\bf QUANTUM GEOMETRODYNAMICS OF THE BIANCHI-IX MODEL}
\vspace*{0.035truein}
\centerline{\bf IN EXTENDED PHASE SPACE}
\vspace*{0.37truein}
\centerline{\footnotesize V.A. SAVCHENKO\footnote{e-mail:
savchenko@phys.rnd.runnet.ru}, T.P. SHESTAKOVA\footnote{e-mail:
stp@phys.rnd.runnet.ru}, G.M. VERESHKOV}
\vspace*{0.015truein}
\centerline{\footnotesize\it Department of theoretical physics, Rostov State
University}
\baselineskip=10pt
\centerline{\footnotesize\it Sorge str. 5, Rostov-on-Don 344090, Russia}
\vspace*{0.225truein}
\pub{17 June 1998}

\vspace*{0.21truein}
\abstracts{ A way of constructing mathematically correct quantum geometrodynamics of
a closed universe is presented. The resulting theory appears to be gauge-
noninvariant and thus consistent with the observation conditions of a closed
universe, by that being considerably distinguished from the traditional Wheeler
-- DeWitt one. For the Bianchi-IX cosmological model it is shown that a
normalizable wave function of the Universe depends on time, allows the
standard probability interpretation and satisfies a gauge-noninvariant
dynamical Schr\"odinger equation. The Wheeler -- DeWitt quantum
geometrodynamics is represented by a singular, BRST-invariant solution to the
Schr\"odinger equation having no property of normalizability.}{}{}

\vspace*{1pt}\textlineskip
\section{Introduction}\label{intro}	
\vspace*{-0.5pt}
\noindent
The homogeneous cosmological Bianchi-IX model is traditionally used as a 
test polygon for various theoretical methods in cosmology. It combines 
mathematical simplicity and physical meaningfulness. The purpose of the 
present work is, making the most of the quantum Bianchi-IX model in 
extended phase space (EPS) as an example, to explore a possibility of 
construction of physically (operationally) interpreted quantum 
geometrodynamics (QGD) by a strict mathematical method without using 
any assumption failing to allow a detailed mathematical proof. As a result of 
our investigation, we have come to the conclusion that mathematically 
correct and physically well-grounded QGD of a closed universe is a 
gauge-noninvariant theory, {\it radically} distinguished from the
Wheeler-DeWitt 
(WDW) QGD.$^1$ by its content.
\pagebreak

\textheight=7.8truein
\setcounter{footnote}{0}
\renewcommand{\thefootnote}{\alph{footnote}}

\section{The many-worlds interpretation of quantum geometrodynamics}

\noindent
The standard QGD is based on the WDW equations
\begin{equation}
\label{WDWeqns}
{\cal T}^{\mu}|\Psi\rangle =0, 
\end{equation}
$$
{\cal T}^0 =
 (-g_{(3)})^{-\frac12}p^{ik}(g_{il}g_{km}-\frac12g_{ik}g_{lm})p^{lm}
 + (-g_{(3)})^{\frac12}R_{(3)}+T^{00}_{(mat)},
$$
$$
{\cal T}^i=-2(\partial_kp^{ik}+\gamma^i_{lm}p^{lm})+T^{0i}_{(mat)},
$$
$p^{ik}$ are the momenta conjugate to the 3-metric $g_{ik}$, 
$\gamma^i_{kl}$ are the three-connections, $g_{(3)}\equiv \det\|g_{ik}\|$, 
$R_{(3)}$ is the 3-curvature, $T^{\mu\nu}_{(mat)}$ is the energy-momentum 
tensor (EMT) of the material fields. Derivation of these 
equations has been discussed by many authors. Because of a number of reasons analyzed 
in details below in Sec.~6. for the Bianchi-IX model, the WDW equations 
are not deducible by correct mathematical methods in the framework of the 
ordinary quantum theory (QT). In principle, this fact itself is not sufficient to 
discard the WDW theory. The ordinary QT is a phenomenological theory for 
describing quasilocal (in a macroscopic sense) phenomena. Therefore its 
extrapolation to the scales of the Universe as a whole is a radical physical 
hypothesis that may be incompatible correctly with the existing formalism. In
this situation it makes sense to analyze the WDW theory as it is, without 
fixing attention on whether a correct way of its construction exists or not.

The most distinctive feature of the WDW theory is that there is no quantum 
evolution of state vector in time. Once adopting the WDW theory, one 
should admit that a wave function satisfying Eqs.~(\ref{WDWeqns}) 
describes the past of the Universe as well as its future with all observers 
being inside the Universe at different stages of its evolution, and all 
observations to be made by these observers. We should emphasize that the 
question about the status of an observer in the WDW theory is rather specific 
since there is no vestige of an observer in Eqs.~(\ref{WDWeqns}) . The 
introduction of the observer into the theory is performed by fixing boundary 
conditions for a universe wave function (WF); we shall return to them below.

First of all, one should pay attention to another feature of the WDW theory: 
because of the status of the universe WF mentioned above this theory does 
not use the postulate about {\it the reduction of a wave packet}. The logical 
coordination of concepts in the WDW theory is carried out within the 
framework of the many-worlds interpretation of the WF proposed by 
Everett.$^2$ and applied to QGD by Wheeler.$^3$. The WF satisfying
Eqs.~(\ref{WDWeqns}) and certain boundary conditions is thought to be a 
branch of a many-worlds wave function that corresponds to a certain 
universe, other branches being selected by other boundary conditions. Thus, 
the boundary conditions for the WF gain a fundamental meaning for the 
theory: they fix all actions of an observer through the whole history of the 
Universe, i.e. they contain the concentrated information about the continuous 
reduction of the Universe WF in the process of evolution of the Universe 
including certain observers inside.$^{2,3,4}$.

Let us discuss the  peculiarities of setting problems in the WDW QGD using 
the Bianchi-IX model as an example, space homogeneity of the latter 
reducing the set of the WDW equations to the only equation
\begin{equation}
\label{B-IX_WDW}
H_{ph}|\Psi\rangle = 0.
\end{equation}
Gauge invariance is expressed by that the choice of a time coordinate is not 
made when deriving (more precisely, when writing down) the WDW
equation. The equation should be solved under some boundary conditions. 
However, carrying out this program one should bear in mind that solutions to 
this equation are unnormalizable. The latter is obvious from the following 
mathematical observations: the WDW equation coincides formally with an 
equation for the eigenfunction of the physical Hamiltonian $H_{ph}$, 
appropriate to the zero eigenvalue. Meanwhile, nothing prevents us from 
studying {\it the whole} spectrum of eigenvalues of the operator $H_{ph}$; 
then the WF satisfying  Eq.~(\ref{B-IX_WDW})turns out to be normalizable 
only if the value E = 0 belongs to a discrete spectrum of the operator 
$H_{ph}$.

As for the operator $H_{ph}$, explicit form of which will be presented in 
Sec.~6. of the paper, it has a {\it continuous} spectrum. In this situation 
one faces the alternative: 1)~to declare the Bianchi-IX model to be meaningless 
and to put the question about searching for such models, whose operator 
$H_{ph}$ has a discrete spectrum line at $E = 0$, or 2) to discard normalizing 
the universe WF, enlarging more, by that, the discrepancy between QGD and 
ordinary QT. In the WDW QGD the second way is chosen that does not 
contradict in principle to the statement about the status of the universe WF 
mentioned above.\fnm{a}\fnt{a}{Strictly speaking, unnormalizability of the
WF means that there is no probabilistic interpretation of it, which is quite 
natural, because the results of  the continuous reduction of a wave packet are 
involved {\it in the boundary conditions} but not in the {\it structure of a 
superposition}. Nevertheless, attempts were made by many authors to 
retrieve a probabilistic interpretation of QGD (see, for example,.$^5$ ) 
analyzing its physical and mathematical contents by methods of the modern 
gauge field theory. In such an approach  mathematical correctness of the 
analysis, thorough study of the question, whether the used mathematical 
procedures do exist, take on fundamental significance. Just this very problem 
will be discussed in details in our paper. The conclusions of our investigation 
are those: mathematically correct procedures exist at least for systems with 
finite number of degrees of freedom, however, such procedures lead to the 
theory radically distinguished from the WDW one by its mathematical form 
as well as by its physical content.}

What should the WDW theory be taken for by an individual local observer? 
Obviously -- for {\it a paradigm} fixing a certain way of thinking that {\it in 
principle} cannot be verified or overthrown experimentally. The reference to 
the fact that in the classical limit of the WDW theory one can obtain the 
Einstein equations, conclusions from which can be compared with
cosmological observations, is not an argument, since it is obvious in advance 
that there exist an {\it infinite number of ways} to make a quantum 
generalization of the classical theory of gravity based not on mathematically 
correct procedures but on adopting some paradigm.

The WDW paradigm may contain a deep sense that is inaccessible for 
understanding yet. However, it is clear that its existence does not deprive 
another approach to QGD problems of a sense; for instance, an approach 
based on adopting another interpretative paradigm or an approach based on 
procedures pretending to a greater extent of mathematical strictness then 
those used in the WDW theory.

\section{Our research algorithm}

\noindent
In our work the following research algorithm is realized.

\begin{romanlist}
\item A path integral (PI) form of a transition amplitude between any two 
states of the Universe is adopted as a basic QGD object (the probabilistic
interpretation of the theory is predetermined by that).
\item Making use of a nonlocal gauge completely removing gauge 
degeneracy, we define an amplitude as a PI not containing divergences  
issuing from infinite number of gauge orbit representatives. 
\item We take notice of the circumstance that a closed universe has no 
asymptotical states in which splitting off the 3-scalar and 3-vector gravitons 
takes place. By this reason we refuse the incorrect operations supposedly 
singling out information about the three-dimensionally transversal modes {\it 
before} finding a universe WF.
\item Instead we state the problem of constructing universe WF containing 
information about a physical object as well as about a reference system (RS), 
fixed by a gauge, in which the object is studied. A gauge is selected by such a 
way that the dynamics of the integrated system ``the physical object  + 
observation means (OM)" is described in EPS. The dynamical Schr\"odinger 
equation (SE) is an unambiguous consequence of the approach; it is derived 
directly from the quantum Hamilton equations in EPS or by the mathematically 
equivalent PI method.
\item The solution of the existence problem of a WF, containing information 
about a physical object as well as about OM, can be obtained in an explicit 
form for the Bianchi-IX model: here we prove that it is possible to
ascertain the structure of the general solution to the gauge-noninvariant 
Schr\"odinger equation (SE).
\end{romanlist}

The results of the realization of this program consist in the following.
\begin{romanlist}
\item The general solution (GS) to the gauge-noninvariant SE is normalizable 
and amenable to the standard Copenhagen interpretation of QT.
\item The transition to WDW theory realized as singling out the BRST-invariant 
sector of SE solutions is not dictated by any general physical or mathematical 
reasons and leads to unnormalizable BRST-invariant wave functions.
\end{romanlist}

\section{The nonlocal gauge and a transition amplitude}

\noindent
The standard quantum theory of gravity is based on the assumption of 
existence of the gauge-invariant transition amplitude
\begin{equation}
\label{GR_ampl}
\langle\,f|i\,\rangle=\frac1{W(g)}
 \int Dg^{\mu\nu}\exp\left(-\frac{i}{2\kappa}\int\sqrt{-g}\,R\,d^4x\right);
\end{equation}
$$
Dg^{\mu\nu}=\prod_x(-g)^\frac52\prod_{\nu\le\mu}\,dg^{\mu\nu}.
$$
$g^{\mu\nu} (x)$ is a 4-metric tensor, $R$ is a scalar curvature, $\kappa$ is the gravity 
constant.

In Eq.~(\ref{GR_ampl}) a normalizing factor $W(g)$ is a divergent integral 
over gauge group space. The numerator of (\ref{GR_ampl}) also contains 
divergences issuing from the infinite number of gauge orbit representatives. 
The ratio of two divergent integrals is supposed to be a finite and physically 
meaningful quantity.

In applications of quantum gravity theory to the quantization problem of the Universe as
a whole (in QGD) the mathematical uncertainty of the amplitude (\ref{GR_ampl}) gains a
fundamental significance. The adduced above statements and generally known 
procedures dealing with the amplitude (\ref{GR_ampl}) are just a heuristic conventions 
concerning carrying out formal operations with the expression which one cannot give a 
definite mathematical sense to. We shall refer to the theoretically unprovable assumption 
about existence of the amplitude (\ref{GR_ampl}) as the hypothesis about a gauge-
invariant regularization ({\it R-hypothesis}).

To make the PI convergent one uses ``the expansion of a gauge unit" followed by 
factoring out the integral over gauge group space. It enables to use the  action extremals 
with gauge-fixing and ghost terms to skeletonize the PI. When carrying out factoring, 
one changes the order of integration in a divergent integral that in principle cannot be 
regularized until the described procedure is over. The assumption that such operations, 
firstly, are admissible and, secondly, do not change properties of the original expression, 
is the essence of the second unprovable {\it F-hypothesis}.

In the above procedure local gauges are used
\begin{equation}
\label{f,h}
f^{\mu}\left(h^{\mu\nu}\right)=0,
\end{equation}
where $h^{\mu\nu}=\sqrt{-g}\,g^{\mu\nu}$, which maintain residual degeneration 
under the diffeomorphism group transformations. The residual transformation 
parameters satisfy the equations
\begin{equation}
\label{M_eta}
\hat{M}^{\nu}_{\mu}\eta^{\mu}=0;
\end{equation}
$$
\hat{M}^{\nu}_{\mu}=\frac{\delta f^{\nu}}{\delta h^{\rho\sigma}}
 \left(-\partial_{\mu} h^{\rho\sigma}
  +\delta^{\rho}_{\mu}h^{\sigma\lambda}\partial_{\lambda}
  +\delta^{\sigma}_{\mu}h^{\rho\lambda}\partial_{\lambda}\right).
$$
The infinitesimal functions
\begin{equation}
\label{delta_h}
\delta h^{\mu\nu}=-\partial_{\lambda}h^{\mu\nu}\eta^{\lambda}
 +h^{\mu\lambda}\partial_{\lambda}\eta^{\nu}
 +h^{\nu\lambda}\partial_{\lambda}\eta^{\mu} 
\end{equation}
describe inertial fields emerging or vanishing under the transformations of 
RS, within the class determined by a local gauge.

We propose a method of constructing the theory without making use of the R- and F-
hypotheses. The method is based on the nonlocal analog of the condition (\ref{f,h}) 
completely removing gauge degeneracy,
\begin{equation}
\label{F,h}
F^{\mu}\left(h^{\lambda\sigma}\right)=
 \int f^{\nu}\left[h^{\lambda\sigma}(x')\right]G^{\mu}_{\nu}(x,x')\,d^4x'=0.
\end{equation}
Here $G^{\mu}_{\nu}(x,x')$ is the retarded Green's function satisfying the
equation
$$
\hat{M}^{\dag\nu}_{\hphantom{\dag}\lambda}G^{\mu}_{\nu}(x,x')=
 \delta^{\mu}_{\lambda}\delta(x-x');
$$
the operator $\hat{M}^{\dag\nu}_{\hphantom{\dag}\lambda}$ is Hermitian 
conjugate to the operator (\ref{M_eta}). The rigidity of the nonlocal gauge condition is 
easily revealed: its variation under infinitesimal transformations of the metric may be 
written as
$$
\mbox{\hbox to\textwidth{$
\displaystyle\delta F^{\mu}=
 \int\left[\frac{\delta f^{\nu}}{\delta h^{\lambda\sigma}}
  \delta h^{\lambda\sigma}(x')\right]G^{\mu}_{\nu}(x,x')\,d^4x'=
 \int\left[\hat{M}^{\nu}_{\lambda}\eta^{\lambda}(x')\right]
  G^{\mu}_{\nu}(x,x')\,d^4x'=
$ \hfil}}
$$
\begin{equation}
\label{delta,F}
 =\int\eta^{\lambda}(x')\hat{M}^{\dag\nu}_{\hphantom{\dag}\lambda}
  G^{\mu}_{\nu}(x,x')\,d^4x'=\eta^{\mu}(x),
\end{equation}
so the equation $\delta F^\mu=0$ has the unique solution $\eta^\mu=0$. 

We propose to adopt the following expression for the transition amplitude
\begin{equation}
\label{amp,t,t0}
\langle\,f,t|i,t_0\,\rangle=
 \int D_Fh^{\mu\nu}\exp\left[i\int\limits_{t_0}^t\!d\tau\!\!\int\!d^3x
  \left(-\frac1{2k}h^{\mu\nu}R_{\mu\nu}\right)\right],
\end{equation}
$$
D_Fh^{\mu\nu}=
 \prod_{\vec{x}}\prod_{t_0<t}M\left[h^{\mu\nu}(\vec{x},\tau)\right]
 \prod_{\nu\le\mu}dh^{\mu\nu}(\vec{x},\tau)
 \prod_{\mu}\delta\left[F^{\mu}(\vec{x},\tau)\right]
$$
as a basic postulate of quantum theory of gravity enabling to avoid the operations with 
divergent PI.

By identity transformations, symbolically written as
$$
\delta\left(F^{\mu}\right)=
 \frac1{\det\|G^{\nu}_{\lambda}\|}\delta\left(f^{\mu}\right)=
 \det\|\hat{M}^{\dag\nu}_{\hphantom{\dag}\lambda}\|\delta\left(f^{\mu}\right),
$$
followed by the representation of the 
$\det\|\hat{M}^{\dag\nu}_{\hphantom{\dag}\lambda}\|$
through a PI over Grassmannian variables $\theta^\mu,\bar\theta_\mu$, the amplitude 
(\ref{amp,t,t0}) is reduced to
\begin{equation}
\label{Damp}
\langle\,f|i\,\rangle=\!
 \int\!D\left(h^{\mu\nu},\pi_{\mu},\theta^{\mu},\bar{\theta}_{\mu}\right)
  \exp\left[i\!\int\!\left(-\frac1{2\kappa}h^{\mu\nu}R_{\mu\nu}
   +\pi_{\mu}f^{\mu}
   +\bar{\theta}_{\nu}\hat{M}_{\mu}^{\nu}\theta^{\mu}\right)d^4x\right];
\end{equation}
\vspace{2pt}
$$
D\left(h^{\mu\nu},\pi_{\mu},\theta^{\mu},\bar{\theta}_{\mu}\right)=
 \prod_x\left(-h\right)^{-\frac52}\prod_{\nu\le\mu}dh^{\mu\nu}
 \prod_{\mu}d\pi_{\mu}\prod_{\nu}d\theta^{\nu}d\bar{\theta}_{\nu}.
$$

The expression (\ref{Damp}) coincides with the known Faddeev -- Popov
amplitude.$^{6,7}$; but, as may be seen from the mathematically correct way it was 
obtained by, there is no ground for a statement that the quantum gravity is a gauge-
invariant theory. Moreover, it is evident, that among the effects contained in 
(\ref{Damp}) gauge-noninvariant effects are present inevitably. As to the question 
whether {\it gauge-invariant} effects are contained in the amplitude, it should be 
analyzed within the theory itself according to properties of the initial and the final states. 
Our investigation shows that in the QGD of a closed universe there exist no gauge-
invariant transition amplitude. 
\pagebreak

\section{The dynamical Schr\"odinger equation for the Bianchi-IX model}

\noindent
The interval in the cosmological model Bianchi-IX looks like.$^8$
$$
ds^2=N^2(t)\,dt^2-\eta_{ab}(t)e^a_ie^b_k\,dx^idx^k;
$$
$$
\eta_{ab}(t)={\rm diag}\left(a^2(t),b^2(t),c^2(t)\right),
$$
$$
\begin{array}{l}
e^1_i=(\sin x^3,-\cos x^3\sin x^1,0),\\
e^2_i=\rule{0pt}{14pt}(\cos x^3,\sin x^3\sin x^1,0),\\
e^3_i=\rule{0pt}{14pt}(0,\cos x^1,1).
\end{array}
$$

We shall also assume that the model includes an arbitrary number $K$ of real
scalar fields described by the Lagrangian
$$
{L_{(scal)}} = \frac1{2\pi^2}\sqrt{-g}\,\left[\frac12\sum _{i=1}^K 
\phi,_{\mu}^{\!\!i}\phi_i^{,\mu}-U_s(\phi_1,\ldots,\phi_K)\right],
$$
and use the following parametrization:
$$
a=\frac12 r\exp\left[\frac12\left(\sqrt{3}\,\varphi+\chi\right)\right];\; b=\frac12 
r\exp\left[\frac12\left(-\sqrt{3}\,\varphi+\chi\right)\right];\;
c=\frac12 r\exp\left(-\chi\right).
$$

Writing out the Einstein equations in the given parametrization it is easy 
to notice that the Bianchi-IX model can be considered as a 
model of a Friedman -- Robertson -- Walker closed universe on which a transversal 
nonlinear gravitational wave $\varphi(t),\chi(t)$ is superposed, $r(t)$ being the scale 
factor. It will be more convenient to use a parameter $q = 2\ln r$.

Denoting
$$Q^a=(q,\varphi,\chi,\phi,\ldots);$$
we shall define the gauge variable $\mu$ through the ``lapse function" $N$  by an 
arbitrary fixed function 
\begin{equation}
\label{zeta}
\zeta(\mu,Q)=\ln\frac{r^3}N
\end{equation}
and confine our attention to the special class of gauges not depending on time
\begin{equation}
\label{mu,f,k}
\mu=f(Q)+k;\ k={\rm const}, 
\end{equation}
or, in a differential form,
$$
\dot{\mu}=f,_a\dot{Q}^a, 
$$
an index after a comma here and further denoting differentiation with respect 
to generalized coordinates: $f,_a=\partial f/\partial Q^a$. Practically, any gauge can be 
represented by Eq.~(\ref{mu,f,k}) using an appropriate parametrization (\ref{zeta}). The 
choice of a differential gauge condition form is dictated by our intention to construct a 
Hamiltonian dynamics in EPS where the Lagrangian multiplier of a gauge in the action 
has to be the momentum canonically conjugate to the gauge variable.

The appropriate effective action, which a PI is based on when quantizing the model, 
reads
\begin{eqnarray}
\label{Seff}
S_{ef\!f}&=&\int dt\biggl\{\frac12\exp\left[\zeta\left(\mu,Q^a\right)\right]
\gamma_{ab}\dot{Q}^a\dot{Q}^b-\exp\left[-\zeta\left(\mu,Q^a\right)\right]
U\left(Q^a\right)\nonumber\\&+&\left(\pi+\dot{\bar{\theta}}\theta
\right)\left(\dot{\mu}-f,_a\dot{Q}^a\right)+\frac i{\zeta,_{\mu}}\dot{\bar
{\theta}}\dot{\theta}\biggr\},
\end{eqnarray}
where $\zeta,_{\mu}=\!\partial \zeta(\mu,Q)/\partial\mu;\;\theta,\bar{\theta}$ are the 
Faddeev -- Popov ghosts after replacement $\bar{\theta}\to-i\bar{\theta}$; indices 
$a,b,\ldots$ are raised and lowered with the ``metric"

$$\gamma_{ab}={\rm diag}(-1,1,1,1,\ldots);$$

$$
U(Q)={\rm e}^{2q}U_g(\varphi,\chi)+{\rm e}^{3q}U_s(\phi),
$$
\begin{eqnarray*}
U_g(\varphi,\chi)&=&\frac23\left\{\exp\left[2\left(\sqrt{3}\,\varphi+
\chi\right)\right]+\exp\left[2\left(-\sqrt{3}\,\varphi+
\chi\right)\right]+\exp(-4\chi)\right.\\&-&2\exp\left[-\left(\sqrt{3}\,
\varphi+\chi\right)\right]-\left.2\exp\left(\sqrt{3}\,\varphi-\chi\right)-
2\exp(2\chi)\right\},
\end{eqnarray*}

The Batalin -- Fradkin -- Vilkovisky (BFV).$^9$ EPS consists of 
the dynamical variables $Q,\mu,\theta,\bar{\theta}$ and the appropriate
canonical momenta
$$
P_a=\frac{\partial L}{\partial\dot{Q}^a},\;\;P_0=\frac{\partial L}{\partial\dot{\mu}}=
\lambda,\;\;\bar{\rho}=\frac{\partial L}{\partial\dot{\theta}},\;\;
\rho=\frac{\partial L}{\partial\dot{\bar{\theta}}}.
$$
The corresponding Hamiltonian
\begin{equation}
\label{Hamilt}
H=\frac12G^{\alpha\beta}P_{\alpha}P_{\beta}-i\zeta,_{\mu}\bar{\rho}\rho+e^{-\zeta}U,
\end{equation}
where $\alpha=(0,a),Q^0=\mu$,
$$
G^{\alpha\beta}={\rm e}^{-\zeta}\left(
\begin{array}{cc}
f,_a\!f^{,a}&f^{,a}\\
f^{,a}&\gamma^{ab}
\end{array}
\right),
$$
gives the canonical equations
\begin{equation}
\label{extrem}
\dot{X}=\left\{H,X\right\},\;\;X=\left(P_{\alpha},\rho,\bar{\rho},Q^{\alpha},
\theta,\bar{\theta}\right). 
\end{equation}

The action (\ref{Seff}) in canonical variables
$$
S=\int\left(\dot{Q}^{\alpha}P_{\alpha}+\dot{\theta}\bar{\rho}+\dot{\bar{
\theta}}\rho-H\right)\,dt 
$$
is invariant, up to dynamically insignificant terms, under BRST and anti-BRST 
transformations in EPS, generated by the charges
\begin{equation}
\label{BRSTgen}
\Omega=H\theta+iP_0\rho,\;\;\bar{\Omega}=H\bar{\theta}-iP_0\bar{\rho}.
\end{equation}
Both the charges are real, nilpotent and mutually anticommutative in the 
sense of Poisson brackets,
$$
\{\Omega,\Omega\}=\{\bar{\Omega},\bar{\Omega}\}=\{\Omega,\bar{\Omega}\}=0,
$$
and are integrals of motion.

The set of equations (\ref{extrem}), which we will refer to as 
{\it conditionally}-classical for the presence of Grassmannian variables in it, is applied to approximate a PI 
by a multiple integral. An important feature of the conditionally-classical model is 
the presence of a specific subsystem, further being referred to as the ``gravitational vacuum 
condensate" (GVC). In (\ref{Seff}) the ``OM Lagrangian" $\lambda(\dot\mu-f,_a\dot 
Q^a)$, where $\lambda=\pi+\dot{\bar\theta}\theta$, corresponds to this subsystem. In a 
simple case, when the gauge function $f(Q^a)$ depends only on $Q^1 = q$ and $\zeta = 
\zeta(\mu,q)$, the quasi-energy-momentum tensor (quasi-EMT) corresponding to this 
Lagrangian is isotropic:
$$
T_{\mu(obs)}^{\nu}={\rm diag}(\varepsilon_{(obs)},-p_{(obs)},-p_{(obs)},-p_{(obs)}),
$$
$$
\varepsilon_{(obs)}=-\frac{\dot{\lambda}}{2\pi^2(\zeta,_{\mu})_k}\exp(\zeta_k-3q),\;\;
p_{(obs)}=\left[1-\frac23\left((\zeta,_q)_k+(\zeta,_{\mu})_kf,_q\right)\right]
\varepsilon,
$$
where an index $k$ denotes that the substitution $\mu=f(Q^a)+k$ has been made. So, the 
GVC is a continuous medium with the equation of state essentially depending on a 
parametrization and a gauge, thus the latter two is proved to be cosmological evolution 
factors. Notice, as well, that investigation of the equations of motion following from the 
effective action (\ref{Seff}) reveals the existence of the conserved quantity
$$
(\zeta,_{\mu})_k^{-1}\dot{\lambda}=E_k. 
$$
thus the latter being a quantitative description of the GVC. As it will be shown below, finding the 
$E_k$ value spectrum is one of the main tasks of gauge-noninvariant QGD.

Proceeding to constructing quantum theory, it is essentially important to note that in the 
EPS formalism a dynamical SE is a direct and unambiguous consequence of canonical 
quantization procedure by no means depending on our concepts about gauge invariance 
or noninvariance of the theory. Really, an SE can be derived from the quantum analogue 
of the canonical equation set (\ref{extrem}) alone:
$$
i\frac{\partial|\Psi\rangle}{\partial t}=H|\Psi\rangle,
$$
where $H$ is the Hamiltonian (\ref{Hamilt}) defined in EPS.

A dynamical SE, surely, can also be obtained in PI formalism having certain 
advantages over operator one. In the latter, as is generally known, the 
operator ordering problem is not resolvable. When deriving an SE from a PI, 
ordering turns to be bound up with a way of a final definition of a PI as a 
limit of a multiple integral, and with a choice of a gauge variable 
parametrization. The parametrization choice determines a PI measure as well 
(the latter being identical with a measure of the WF normalizing integral -- 
probability measure). Let us consider a PI for a WF in the coordinate 
representation. Such a WF, according to the stated above, is defined on the 
extended configurational space with the coordinates $Q^a,\mu,\theta,
\bar{\theta}$:
\begin{eqnarray}
\label{WF-PI}
{\Psi(Q^a,\mu,\theta,\bar{\theta};t)}&=&\int\langle Q^a,\mu,\theta,\bar{\theta};
t|Q^a_{(0)},\mu_{(0)},\theta_{(0)},\bar{\theta}_{(0)};t_0\rangle
\Psi(Q^a_{(0)},\mu_{(0)},\theta_{(0)},\bar{\theta}_{(0)};t_0)\nonumber\\
&\times& M\left(Q^a_{(0)},\mu_{(0)}\right)\,d\theta_{(0)}d\bar{\theta}_{(0)}
d\mu_{(0)}\prod_b dQ^b_{(0)}.
\end{eqnarray}
The transition amplitude, appearing in (\ref{WF-PI}),
\begin{eqnarray*}
\lefteqn{\langle Q^a,\mu,\theta,\bar{\theta};t|Q^a_{(0)},\mu_{(0)},\theta_{(0)},
\bar{\theta}_{(0)};t_0\rangle=}\\
&& C\int\exp[iS(t,t_0)]\prod_{t_0<\tau<t}M\left(Q^a_{(\tau)},\mu_{(\tau)}
\right)\,d\mu_{(\tau)}d\theta_{(\tau)}d\bar{\theta}_{(\tau)}
\prod_bdQ^b_{(\tau)}d\lambda_{(\tau)}d\lambda_{(t)},
\end{eqnarray*}
is given by the gauged action (\ref{Seff}).

It  is well known that a PI is not defined in internal terms. Proceeding from the 
standard treating, we shall consider it as a limit at $\epsilon_i\to0$ of the following 
integral:
\begin{eqnarray*}
\Psi^{(N)}(Q^a,\mu,\theta,\bar{\theta})&=&C\int\exp\left\{i\sum^N_{i=1} 
S(t_i,t_{i-1})\right\}\Psi^{(0)}(Q^a,\mu,\theta,\bar{\theta})\\
&\times&\prod^{N-1}_{i=0}M\left(Q^a_{(i)},\mu_{(i)}\right)\,
d\mu_{(i)}d\theta_{(i)}d\bar{\theta}_{(i)}\prod_bdQ^b_{(i)}d\lambda_{(i+1)},
\end{eqnarray*}
where $t_i-t_{i-1}=\epsilon_i$,
\begin{eqnarray*}
S(t_i,t_{i-1})&\approx &\epsilon_i\left\{\frac12\exp(\zeta_{(i)})
\dot{Q}^a_{(i)}\dot{Q}_a^{(i)}-\exp(-\zeta_{(i)})U(Q^a_{(i)})\right.\\
&+&\left.\lambda_{(i)}\left[\dot{\mu}_{(i)}-\dot{f}(Q^a_{(i)})\right]+
\frac i{\zeta,_{\mu(i)}}\dot{\bar{\theta}}_{(i)}\dot{\theta}_{(i)}\right\}.
\end{eqnarray*}

Operating by the standard Feynman method (see.$^{10}$), in the first order 
one obtains the SE
\begin{equation}
 \label{SE1}
i\frac{\partial\Psi(Q^a,\mu,\theta,\bar{\theta};t)}{\partial t}=H\Psi(Q^a,
\mu,\theta,\bar{\theta};t)
\end{equation}
the zero-order terms give the constraint
between the measure $M$, step $\epsilon$ and parametrization function
$\zeta$:
\begin{equation}
\label{eps,M-coupl}
\frac1{\epsilon \zeta,_{\mu}}\left(\epsilon {\rm e}^{-\zeta}\right)^
{\frac K2}M={\rm const}.
\end{equation}
The requirement of the Hamiltonian to be Hermitian gives rise to another
constraint between the measure and the parametrization,
\begin{equation}
\label{zeta,M-coupl}
M={\rm const}\cdot \zeta,_{\mu}\exp\left(\frac{K+3}2\zeta\right).
\end{equation}
The independence of the parameter $\epsilon$ on the variables $Q^a,\mu$
follows from(\ref{eps,M-coupl}), (\ref{zeta,M-coupl}). 

The Hamiltonian in the SE obtained by the PI method is presentable in the form
$$
H=- i\zeta,_{\mu}\frac{\partial}{\partial\theta}\frac{\partial}{\partial\bar{\theta}}- 
\frac1{2M}\frac{\partial}{\partial Q^{\alpha}}\tilde G^{\alpha\beta}
\frac{\partial}{\partial Q^{\beta}}+{\rm e}^{-\zeta}(U-V), 
$$
where $M$ is defined by the formula (\ref{zeta,M-coupl}), 
$\tilde G^{\alpha\beta}=MG^{\alpha\beta}$,

\begin{eqnarray}
\label{V}
V=&-&\frac3{12}\frac{(\zeta,_{\mu})^a(\zeta,_{\mu})_a}{\zeta,_{\mu}\!\!\!^2}+
\frac{(\zeta,_{\mu})^a_a}{3\zeta,_{\mu}}+\frac{K+1}{6\zeta,_{\mu}}\zeta_a 
(\zeta,_{\mu})^a\nonumber\\&+&\frac1{24}(K^2+3K+14)\zeta_a\zeta^a 
+\frac{K+2}6\zeta_a^a, 
\end{eqnarray}
$\zeta_a=\partial \zeta/\partial Q^a+f,_a\partial \zeta/\partial\mu$.

The general solution (GS) to the SE in the coordinate representation is a WF 
$$\Psi=\Psi(t,Q^a,Q^0,\theta,\bar{\theta}),$$ depending on time $t$, physical 
variables $Q^a$, gauge variable $Q^0\equiv\mu$, and ghost variables
$\theta,\bar{\theta}$. Making use of the Hamiltonian structure only, one can ascertain 
the WF dependence on the variables $Q^0,\theta,\bar{\theta}$.

To begin with, note that in the class of gauges (\ref{mu,f,k}), not depending 
on time explicitly, the GS to the SE (\ref{SE1}) is expandable in the 
stationary state eigenfunctions satisfying the stationary SE
\begin{equation}
\label{station_SE}
H\Psi_n(Q^a,Q^0,\theta,\bar{\theta})=E_n\Psi_n(Q^a,Q^0,\theta,\bar{\theta}).
\end{equation}
One of the canonical equations, the gauge equation
$$\left[H,Q^0-f(Q^a)\right]=0$$
means the commutativeness of the Hamiltonian with the operator $Q^0-f(Q^a)$ and, 
consequently, an arbitrary solution to Eq.~(\ref{station_SE}) can be presented 
in the form of a superposition of this operator eigenstates $|k\rangle$,
$$\{Q^0-f(Q^a)\}|k\rangle=k|k\rangle.$$
The same concerns the GS to (\ref{SE1}) as a superposition of the stationary
states. In the coordinate representation
\begin{equation}
\label{k-vector}
|k\rangle=\delta(Q^0-f(Q^a)-k), 
\end{equation}
so the GS to the SE has the structure
\begin{equation}
\label{delta-factor}
{\Psi(Q^a,Q^0,\theta,\bar{\theta};t)}=\int \Phi_k(Q^a,\theta,
\bar{\theta};t)\delta(Q^0-f(Q^a)-k)\,dk.
\end{equation}
So far as there is no other (independent) integral of motion for the $Q^0$ 
variable, the functions (\ref{k-vector}) make the only basis depending on the 
gauge variable $Q^0$, i.e. the GS to the dynamical SE inevitably has the 
structure (\ref{delta-factor}). One can come to the same conclusion by 
making detailed analysis of the WF in the PI formalism. 

The SE for $\Phi_k(Q^a,\theta,\bar{\theta};t)$ reads
\begin{equation}
\label{Phi_k-SE}
i\frac{\partial \Phi_k(Q^a,\theta,\bar{\theta};t)}{\partial t}=
H_k\Phi_k(Q^a,\theta,\bar{\theta};t),
\end{equation}
$$
H_k=- i(\zeta,_{\mu})_k\frac{\partial}{\partial\theta}\frac{\partial}
{\partial\bar{\theta}}-\frac12\exp(-\zeta_k)\left(\frac{\partial^2}
{\partial Q^a\partial Q_a}+Z^a_k\frac{\partial}{\partial Q^a}\right)+
\exp(-\zeta_k)(U-V),
$$
$$
Z^a_k=\frac{(\zeta,_{\mu})^a_k}{(\zeta,_{\mu})_k}+\frac{K+1}2\zeta^a_k,
$$
$V$ being defined by Eq.~(\ref{V}), $(\zeta^a)_k=\zeta_k^{,a}\equiv\partial 
\zeta_k/\partial Q_a$.

So, the WF dependence on $\mu=Q^0$ is determined by Eq.~(\ref{delta-factor}). As 
will be shown below, such a structure of the WF, on a certain restriction on 
the $\Phi_k$ dependence on $k$, makes the normalizing integral over the variable 
$Q^0$, transformed into the integral over $k$, to be convergent. As for the 
dependence on the ghosts, it is strictly enough fixed by the SE in combination
with the usual demand of the norm positivity. Indeed, in the general case the 
WF can be presented by the series in the Grassmannian variables,
$$
\Phi_k(Q^a,\theta,\bar{\theta};t)=\Psi^0_k(Q^a,t)+\Psi^1_k(Q^a,t)\theta
+\Psi^{\bar{1}}_k(Q^a,t)\bar{\theta}+\Psi^2_k(Q^a,t)\bar{\theta}
\theta.
$$
After substitution into (\ref{Phi_k-SE}) one obtains the equations for the 
components
\begin{equation}
\label{ghost-expan-SE1}
i\frac{\partial\Psi^0_k}{\partial t}=H^0_k\Psi^0_k-i(\zeta,_{\mu})_k\Psi^2_k,
\end{equation}
\begin{equation}
\label{ghost-expan-SE2}
i\frac{\partial\Psi^i_k}{\partial t}=H^0_k\Psi^i_k,\;\;i=1,\bar{1},2,
\end{equation}
$$
H^0_k=H_k+i(\zeta,_{\mu})_k\frac{\partial}{\partial\theta}\frac{\partial}{\partial 
\bar{\theta}},
$$
and the normalization condition puts the constraints on these components: from 
the norm positivity demand it follows that
$$
- i\int\left(\Psi^{0*}_k\Psi^2_k-\Psi^{2*}_k\Psi^0_k+
\Psi^{\bar{1}*}_k\Psi^1_k-\Psi^{1*}_k\Psi^{\bar{1}}_k
\right)\bar{\theta}\theta\, d\theta d\bar{\theta}>0.
$$
One of the consequences of the inequality is $\Psi_k^2=0$, or $\Psi_k^0=0$, 
or $\Psi_k^2=i\Psi_k^0$, and Eqs.~(\ref{ghost-expan-SE1}), 
(\ref{ghost-expan-SE2}) reduce all the three versions to the one,
$$\Psi^0_k=\Psi^2_k=0,$$
$$\Psi^1_k=i\Psi^{\bar{1}}_k,$$
so, finally,
\begin{equation}
\label{Phi-of-ghost}
\Phi_k(Q^a,\theta,\bar{\theta};t)=\Psi_k(Q^a,t)(\bar{\theta}+i\theta),
\end{equation}
where $\Psi_k(Q^a,t)$ is a solution to the equation
\begin{eqnarray*}
i\frac{\partial\Psi_k(Q^a,t)}{\partial t}&=&
-\frac1{2M_k}\frac{\partial}{\partial Q^a}M_k
\exp(-\zeta_k)\gamma^{ab}\frac{\partial\Psi_k(Q^a,t)}{\partial Q^b}\nonumber\\ 
&+&\exp(-\zeta_k)(U-V_k)\Psi_k(Q^a,t),
\end{eqnarray*}

$$
M_k=(\zeta,_{\mu})_k\exp\left(\frac{K+3}2\zeta_k\right),
$$
The unitarity property of a physical state WF takes the form
$$
\int\Psi^*_{k'}(Q^a,t)\delta(\mu-f(Q^a)-k')\Psi_k(Q^a,t)
\delta(\mu-f(Q^a)-k)\,dkdk'M(Q^a,\mu)\,d\mu\prod_{a=1}^{K+3}dQ^a
$$
\begin{equation}
\label{norm.int}
=\int\Psi^*_k(Q^a,t)\Psi_k(Q^a,t)\,dk\prod_adQ^a.
\end{equation}

Thus, the GS (\ref{delta-factor}), (\ref{Phi-of-ghost}) to Eq.~(\ref{SE1}),
under the condition the $\Psi_k(Q^a,t)$ to be a sufficiently narrow packet 
over $k$, is normalizable with respect to the gauge variable, as well as to 
the ghosts and the physical variables.

The peculiarity of the amplitude (\ref{Phi-of-ghost}) lies in the fact that 
the theory does not control its dependence on the free parameter $k$. From
the standpoint of the classical dynamic equations the parameter $k$ sets an
initial condition for the variable $\mu$ and, by the same, determines an 
initial clock setting. In QT, however, there exist no physical state with a fixed $k$ value. 
Really, the unitarity requirement (see (\ref{norm.int})) allows the existence of a physical 
state represented by a packet over $k$, narrow enough to fit a certain classical $\bar{k}$ 
value, but not by a $\delta$-shaped packet. So, in the theory appears an additional degree 
of freedom that could be named {\em an observer degree of freedom}. Unlike the 
quantum uncertainties associated with operator noncommutativeness, QGD do not 
control even width of a $k$ packet.

The GS structure
\begin{equation}
\label{time-depend.WF}
\Psi(Q^a,Q^0,\theta,\bar{\theta};t)=\int\Psi(E_k;Q^a)\exp(-iE_kt)
(\bar{\theta}+i\theta)\delta(\mu-f(Q^a)-k)\,dE_kdk,
\end{equation}
where $\Psi(E_k;Q^a)$ is a solution to the stationary equation
$$
H^0_k\Psi_k(Q^a)=E_k\Psi_k(Q^a),
$$
shows that the WF carries information on 1) a physical object, 2) OM, 3) correlations 
between a physical object and OM. OM are represented by the factored part of the WF -- 
by the $\delta$-function of a gauge and by the ghosts; a physical object is described 
by the function $\Psi_k$; correlations are present in the effective potential 
$V_k$, which the solution depends on, an also in the WF time dependence, or 
after going over to the stationary states, they are present in the effective 
potential and the spectrum $E_k$.

It is of principal significance to emphasize that the construction procedure 
of the WF (\ref{time-depend.WF}) is the only strict mathematical method to do 
it, by no way corresponding to the QGD WDW. The question arise: do 
Eq.~(\ref{SE1}) and its solutions have any relation to the WDW theory? In other 
words, whether it is possible to pick out such a physical part of the 
structure (\ref{time-depend.WF}) that would satisfy the WDW equation and could
be reasonably interpreted?

\section{Mathematical and physical problems of the gauge-invariant QGD}

\noindent
To begin with, let us discuss a possibility of constructing a WF, 
corresponding to QGD WDW, by going over from the GS to some particular 
solution. As it is known, the transition to QGD WDW entails the separation of the 
physical variable subspace from EPS. For this purpose it is not enough to 
separate the physical part $\Psi_k(Q^a)$ from the GS: one also has to banish 
correlations between the physical object properties and those of OM. The 
latter are given, first of all, by the GVC parameter. Hence, to eliminate 
correlations, firstly, we put $E_k=0$. Secondly, making use of the noted in 
Sec.~5. possibility of going over to any given gauge function 
$f(Q^a)$ by means of the transformation of the parametrization function $\zeta(\mu,Q^a)$,  
it is necessary to pass to the  gauge
\begin{equation} \label{triv.gauge}\mu=k. \end{equation}
And, thirdly, the measure should be factored: $M=M_1(\mu)M_2(Q^a)$. In view of
Eq.~(\ref{zeta,M-coupl}), the measure factorization, in turn, requires factoring 
the parametrization that means
\begin{equation}
\label{zeta_add}
 \zeta(\mu,Q^a)=\zeta_1(\mu)+\zeta_2(Q^a), 
\end{equation}
and conservation of additivity of the function (\ref{zeta_add}) when passing to 
the gauge (\ref{triv.gauge}) puts one more restriction on the parametrization 
function: $\zeta_1$ should be linear in $\mu$,
\begin{equation}
\label{zeta1_lin}
 \zeta_1(\mu)=A+B\mu. 
\end{equation}
Under these conditions one obtains the WDW equation for the physical part of 
the WF
\begin{equation}
\label{WDWeqn}
 H_{ph}\Psi(Q^a)=0, 
\end{equation}
\begin{equation}
\label{H_ph}
H_{ph}=-\frac1{2M_2}\frac{\partial}{\partial Q^a}M_2\exp(-\zeta_2)\gamma^{ab}
\frac{\partial}{\partial Q^b}+\exp(-\zeta_2)U+\frac16{\bf R}, 
\end{equation}
\begin{equation}
\label{R}
{\bf R}=-\exp(-\zeta_2)\left[\frac14\left(K^2+3K+14\right)\zeta_2^{,a}\zeta_2,_a
+\left(K+2\right)\zeta_2^{,a},_a\right] 
\end{equation}
being the scalar curvature constructed of the metric $G_{ab}=\exp(\zeta_2)
\gamma_{ab}$.

Eq.~(\ref{WDWeqn}) possesses all formal properties of the WDW equation
including parametrization noninvariance and the lack of any visible vestige of a
gauge. {it However, the described above method of deriving  Eq. (\ref{WDWeqn}) makes
it apparent that in the gauge class (\ref{mu,f,k}), any change of the parametrization
function $\zeta(\mu,Q^a)$ is mathematically equivalent to a new gauge. Hence, the generally known
parametrization noninvariance of the WDW theory, as a matter of fact, is the ill-hidden
gauge noninvariance}. This circumstance has to be taken into account when estimating
the status of the WDW theory.

On the other hand, by origin, the single Hamiltonian eigenvalue $E_k=0$ fixed by
the equation represents the {\em line in the continuous spectrum} of the
Hamiltonian (\ref{H_ph}), hence the solution is unnormalizable. In other words,
on that way of formal singling out the particular solution to Eq.~(\ref{SE1})
one would fail to obtain a WF having physical meaning generally adopted in
QT.\fnm{b}\fnt{b}{A positive solution to this problem would need, apart from a
discrete $H_{ph}$ spectrum, the presence of the line $E=0$ in it.}

The other approach to the existence problem of gauge invariant states is based
on picking out the {\em singular}, BRST-invariant solutions. The
BFV scheme assigns the central part of maintaining gauge invariance to the
BRST generators (\ref{BRSTgen}): state vectors are primordially obey
the superselection rules
\begin{equation}
\label{BRST-constr}
\Omega|\Psi\rangle=0,\;\;\bar{\Omega}|\Psi\rangle=0.
\end{equation}
These equations signify BRST and anti-BRST invariance of the quantum states,
the BRST-equivalence classes of state vectors may be represented by
vectors independent of ghosts and gauge degrees of freedom.$^9$.

Notice at once that the conditions (\ref{BRST-constr}) are to be treated as an
independent postulate: BRST invariance of the action cannot serve as a reason
for them. It is known that an action invariance under some global
transformations does not mean the state vector invariance; state vectors just
have to be {\em covariant}, subject to the appropriate unitary
transformations.

The fact that a BRST-invariant WF is unobtainable from the GS can be seen
by the following: in view of the formulae (\ref{BRSTgen}) for the generators
$\Omega,\bar{\Omega}$, the strictly defined GS (\ref{time-depend.WF})
dependence on the ghosts and the variable $\mu$ makes impossible to put the
conditions (\ref{BRST-constr}) on the GS. The BRST-invariant WF represents a
{\em singular} solution, the quantum analogue of the mentioned above
{\em degenerate} Cauchy problem. Hence, the question arises about its
physical meaning, and {\em normalizability} again serves as a criterion.

For the functions independent of the ghosts the conditions (\ref{BRST-constr})
come to the equation
\begin{equation}
\label{H_0-eqn}
 H_0\Psi(Q^{\alpha})=0,
\end{equation}
where
\begin{equation}
\label{H_0-def}
H_0=-\frac1{2M}\frac{\partial}{\partial Q^{\alpha}}\tilde{G}^{\alpha\beta}
\frac{\partial}{\partial Q^{\beta}}+{\rm e}^{-\zeta}(U-V).
\end{equation}
This equation corresponds to the classical constraint
$$H_0^{cl}=-\frac12 G^{\alpha\beta}P_{\alpha}P_{\beta}+{\rm e}^{-\zeta}U=0.$$
The physical part is singled out from the WF $\Psi(Q^a,Q^0)$ with the help of
the additional condition in the form of the quantum version of the other
classical constraint,
$$P_0=0,$$
Taking account of the form of a self-conjugate momentum operator in the presence of  a
nontrivial measure,
$$
P_{\alpha}=-i\left(\frac{\partial}{\partial Q_{\alpha}}+
\frac{M,_{\alpha}}{2M}\right),
$$
the condition is represented by the equation
\begin{equation}
\label{P_0-constr}
\frac{\partial\Psi}{\partial Q^0}=-\frac1{2M}\frac{\partial M}{\partial Q^0}\Psi.
\end{equation}
After factoring the measure and transforming the parametrization function,
required when going over to the QGD WDW (see (\ref{triv.gauge}) --
(\ref{zeta1_lin})), the solution to Eq.~(\ref{P_0-constr}) takes the form
\begin{equation}
\label{Psi-M-fact}
\Psi(Q^a,Q^0)=\Psi(Q^a)M_1^{-\frac12}(Q^0),
\end{equation}
and Eq.~(\ref{WDWeqn}) -- (\ref{R}) for the WF $\Psi(Q^a)$ (\ref{Psi-M-fact})
follows from Eq.~(\ref{H_0-eqn}) -- (\ref{H_0-def}). However, this time the
derivation procedure of the equation does not consist in fixing a value in a
Hamiltonian spectrum, but in {\em reducing} the spectrum {\em itself} to the
single value by means of the conditions (\ref{BRST-constr}) (confining the
Hamiltonian space of definition to the BRST-invariant vectors).

Nevertheless, the WF normalizability condition is not satisfied here either.
In the EPS formalism a WF norm, primordially defined in the extended
configurational space, contains the integral over the $\mu$ variable,
divergent in the case of BRST-invariant WF. On the other hand, the integral
over the ghosts vanishes, and in the standard BFV scheme the existence of an
interim regularization of the normalizing integral is supposed that would
result in a finite, in particular, unity norm. However, the possibility of a
mathematically satisfactory interim regularization giving the $c$-valued
unity made of $c$-valued infinity combined with the zero {\em Grassmannian}
integral, is not obvious, in any case, it needs additional assumptions not
present within the theory.

Hence, the singular BRST-invariant solutions do not satisfy mathematical
correctness criteria either. Therefore the GS structure
(\ref{time-depend.WF}) presents the only possible correct WF structure in
the consistent QGD, all the other solutions failing to satisfy the criteria of
mathematical correctness and physical rationality.

There is still the third way to obtain WDW Eq.~(\ref{WDWeqn}) -- (\ref{R})
on which the question about normalization
does not arise directly, or more precisely, it does not get such a definite
general answer. Without adducing appropriate calculations we will note that
on this way two incorrect mathematical structures are necessarily used:
\begin{itemlist}
\item a gauge-invariant PI on partially degenerate action extremals without
pointing out a procedure of removing coordinate effects;
\item a WF definition through a divergent integral over gauge variables.
\end{itemlist}
But realization of this program necessarily requires to use, as its component, a procedure
of eliminating coordinate effects, described by the expression (\ref{delta_h}). Availability
of such a procedure is provided either by existence of local canonical gauges, completely
removing degeneracy, or by dynamical splitting off the ``nonphysical" degrees of freedom
from the physical ones in asymptotical states. In quantum theory of gravity, as it is
known, there is no gauge with the mentioned property, and in QGD of a closed universe
there exist no asymptotical state enabling a dynamical splitting operation. So, the third
way to derive the WDW equation, that is met with in publications (see, e.g.$^5$), cannot
be recognized to be correct, as well, for it ignores a procedure of removing coordinate
effects.

The described methods to derive the WDW equation reveal the following.
\begin{romanlist}
\item The WF not containing information about correlations between the
physical object and OM is the same in all the approaches.
\item If the state vector allows to compute average values of observable quantities, then
information about the correlations has to be contained in it inevitably. In other words,
there is no {\em physical} (normalizable) quantum state without a GVC.
\end{romanlist}

It is to be ascertained that there is no QGD as a gauge-invariant theory of
physical states in a closed universe based on the general grounds of
QT. The illusion of the existence of such a theory would arise if one forgets about
the necessity of singling out gauge orbit representatives and tries to come to
agreement about some special quantization rules.

As it was already mentioned in Sec.~2., the WDW theory supplemented
with the many-worlds interpretation of a WF, according to its status of a global quantum
theory, represents a new paradigm, but not a consequence of the existing quantum theory.
Hence, the question is not whether it can be obtained correctly (we have seen above that
one cannot do it), the question is, what formal grounds do we have for the introduction of the
new paradigm? Usually the principle of gauge invariance of the theory is adduced as such
grounds. However, we should emphasize that these grounds do not exist either: as it was
shown above, the parametrization-noninvariant WDW equation (32) is, in fact,
gauge-noninvariant.

\section{On the physical interpretation of the gauge-noninvariant QGD}

\noindent
As to the correct PI method, it shows unambiguously that, being applied to
a closed universe, the ordinary QT of
gravity is gauge-noninvariant. And this feature of the theory is the evidence of
its {\em adequacy} to the phenomena in question: it answers to the
conditions of observations of a closed universe, where there is no possibility
to remove an instrument for infinite distance from the object and thus to
avoid influence of inertial fields, locally indistinguishable from
gravitational field, on the instrument.

As it was shown by Landau and Lifshitz.$^8$, full information about dynamical
geometry can be obtained directly in experimental way (without theoretical
reconstruction) only in an RS disposed on a medium filling the whole space.
Thus the medium have to be {\it continuous}. The choice of a certain RS within this class is fixed by a
gauge condition that means breaking the space-time symmetry (diffeomorphism group
symmetry) by the medium. The operational interpretation of a GVC as such a medium, a
carrier of a Landau -- Lifshitz RS, follows from the foregoing.

Appealing to Landau -- Lifshitz RS makes the statement about the gauge noninvariance
of quantum theory of gravity be almost obvious. Indeed, a formal transformation of
coordinates meaning a transition to another gauge, physically corresponds to removing
OM from {\it the whole space} of the Universe and replacing them by other OM. From
the viewpoint of the standard Copenhagen interpretation of QT that declares existence of
unremovable connections between the properties of an object and OM, it seems to be
incredible that such an operation performed on the {\it whole Universe} scale, would not
result in changing its quantum properties.\fnm{c}\fnt{c}{ In the theory of quantum
transitions between asymptotic states it is obvious that the replacement of bodies on
which an asymptotically inertial RS is realized and replacement of detectors disposed on
these bodies cannot change a physical situation. Therefore such a theory {\it has} to be
gauge-invariant.}

\nonumsection{References}

\end{document}